\documentstyle[epsf,12pt]{article}
\newcommand {\eq}{\begin{equation}}
\newcommand {\qe}{\end{equation}}

\newcommand {\bfr}{{\bf r}}
\newcommand {\bfq}{{\bf q}}
\newcommand {\bfs}{{\bf s}}
\newcommand {\bfR}{{\bf R}}
\newcommand {\ea} {{\it et al.}}

\newcommand {\bfk}{{\bf k}}
\newcommand {\bfp}{{\bf p}}
\newcommand {\prc}{Phys. Rev. C}
\newcommand {\h}{\frac{1}{2}}

\newcommand {\pim} {$\pi^-$}
\newcommand {\pip} {$\pi^+$}
\newcommand {\ph} {\phi}
\newcommand {\pr}{{\it Phys. Rev. }}
\newcommand {\prl}{{\it Phys. Rev. Lett. }}
\newcommand {\pl}{{\it Phys. Lett. }}
\newcommand {\nucp}{{\it Nucl. Phys. }}
\newcommand {\bfP}{{\bf P}}
\newcommand {\ppi}{{\bf p}_{\pi}}

\begin{document}
\baselineskip=1.2\baselineskip

\begin{center}                                   
{\Huge Pion double charge exchange on $^4$He}

\vspace{.5in}

{\large M. Alqadi and W. R. Gibbs}

\vspace{.15in}

Department of Physics, New Mexico State University, Las Cruces, NM,
88003\\
\today
\end{center}

\vspace{.5in}     
  
\abstract{ The doubly differential cross sections for the
$^4$He$(\pi^+,\pi^-) 4p$ reaction were calculated using both a two-nucleon
sequential single charge exchange model and an intranuclear cascade code.  
Final state interactions between the two final protons which were the
initial neutrons were included in both methods.  At incident pion energies
of 240 and 270 MeV the low-energy peak observed experimentally in the
energy spectrum of the final pions can be understood only if the
contribution of pion production is included.  The calculated cross
sections are compared with data. }

\newpage

\section{Introduction}  
                                                     
The pion double-charge exchange (DCX) reaction is one of the rare forms of
nuclear reactions which involve a minimum of two nucleons and provides a
tool to study two nucleon correlations within a target nucleus. A number
of studies have been made of exclusive DCX and interesting results have
been obtained in regard to correlations\cite{corrs} and the propagation of
pions through the nucleus in its ground state\cite{mutazz,wu}

In this paper we study inclusive DCX on a well understood
nucleus where the principal physics interest lies in 
understanding and describing the mechanism.

It is essential to be able to describe nuclear reactions on the hadronic
level in order to interpret possible deviations from the expected behavior
as ``new physics'' phenomena. Several techniques have been used to
describe inclusive DCX. Becker and Schmit \cite{BS} used the sequential
single charge mechanism in their calculations and their results showed
general agreement with the energy dependence of the total cross section
but predicted cross sections larger than the experiment data.

Germond and Wilkin\cite{GW} suggested a different mechanism for DCX
reaction, the interaction of the incident pion with the exchange current
meson or pions in the cloud surrounding each nucleon. Their calculations
did not include the Pauli effect or Fermi motion of the nucleons.
The production mechanism introduced by Jeanneret \cite {jeanneret} gave
good agreement with DCX cross section at energies above the P$_{33}$ 
region and may be the main mechanism there.

Gibbs \ea \cite {Gibbs1}, reported another attempt to calculate the total
and doubly differential cross section for the DCX reaction by using the
sequential single charge exchange mechanism. They included a 
treatment of anti-symmetric wave functions. Both Pauli blocking and a
crude form of final state interaction (FSI) were included in their work
but interactions with the ``spectator'' protons were ignored as in the 
previous cases.

In conjunction with dibaryon searches, Gr\"ater \ea \cite{Bilger}
calculated the total cross section for $^4$He. In their model FSI was
taken into account by using a Watson-Migdal Model in an eikonal
approximation.

Following the early total DCX cross section data, extensive experimental
studies performed at Clinton P. Anderson Meson Physics facility reported
high precision results for doubly differential cross sections. The
experiments covered a wide range of elements (A=4, 208) over a range of
incident energies (120 - 270) MeV. These studies showed a two peak
structure in the energy spectra at forward angles at the higher energies
in the low atomic number elements such as $^4$He and $^3$He \cite
{kinney,Yuly}, while that structure disappears in the heavier elements
$^{16}$O, $^{40}$Ca \cite{wood,wood1} .

A number of calculations attempting to explain this two-peaked structure
have followed a suggestion by M. Thies.  He and Van Loon \cite{thies} did
calculations based on the fact that the angular distribution of the pion
nucleon cross section is dominated by the p-wave and hence has a
forward-backward peaked angular distribution in the $\pi$N center of mass.  
In this case, pions coming from the DCX reaction in the forward direction
would have made two small angle or two large angle scatterings.  Two small
angle scatterings would generate only a small energy loss while two large
angle scatterings would bring about a large energy loss due to the recoil
of the nucleons. This effect surely exists but there have been
difficulties with obtaining a quantitative representation of the data
using only this model.

Another early attempt to explain the double peak structure in $^4$He was
by Wood\cite {wood} relying on the mechanism proposed by Thies. Kinney
\cite{kinney} modified the Thies model by including several effects of the
nuclear medium. His results gave reasonable quantitative agreement with
the shape of the spectrum for $^4$He for some angles, but did not follow
the cross section as a function of angle. Recent calculations carried out
by Kulkarni\cite {kul}, based on a relativized version of the code of
Ref.\cite{Gibbs1}, found qualitative agreement with Kinney's data.

Part of the difficulty with the forward-backward argument is that the
scatterings take place over a 3 dimensional sphere, which means that there
is factor of $\sin\theta$ arising from the solid angle of the first charge
exchange which mitigates the forward-backward angular distribution.  
Also, at 240 and 270 MeV, where the effect is the strongest, not only does
the pion-nucleon angular distribution become more forward peaked in the
center of mass as energy increases but it will also be thrown forward in
the laboratory so that the backward (corresponding to low energy) peak is
much diminished.  We will return to this point in the discussion.

\section{Two-nucleon Model}

We first consider a model in which the two ``spectator'' protons do not
play an active role in the scattering but do influence the result through
the energy that they can carry off.

\subsection{Calculation of Matrix element}
        
We assume a sequential mechanism to describe the DCX reaction in
$^4$He$(\pi^+,\pi^-) 4p$. The operator for the pion DCX amplitude in the
plane wave limit \cite {Gibbs2} can be written as

\eq
F_{D}(\bfk,\bfk',\bfr_{1},\bfr_2)=\frac{1}{2\pi^2} \int d\bfq
e^{-i\bfk'\cdot\bf r_2}f_2(\bfq,\bfk') 
\frac{e^{i\bfq\cdot(\bfr_2-\bfr_1)}}{q^2-k_0^2} 
f_1(\bfk,\bfq) e^{i\bfk\cdot\bf  r_1},
\qe
where the pion-nucleon single charge exchange amplitude is represented
as
\eq
f(\bfk,\bfq) = [\lambda_0+\lambda_{1} \bfk\cdot\bfq+i \lambda _{f}
\sigma_2 \cdot(\bfk \times \bfq) ] v(q)v(k).
\qe 
Here $\frac{1}{q^2-k_0^2}$ is the pion propagator and v(q) is the
off-shell form factor taken here to be,
\eq
 v(q)=\frac{\Lambda^2+k^2}{\Lambda^2+q^2}. 
\qe

The quantity $\Lambda$ (assumed here to be 4 fm$^{-1}$) is related to the
range of the pion-nucleon interaction and the $\lambda $'s are obtained
from $\pi$-N phase shifts\cite{arndt}.  $\bfr_1 $ and $\bfr_2 $ denote the
position vectors of the members of the struck nucleon pair. We first
consider the double scattering to occur from two neutrons leading to a
definite final state of 4 protons, as if it were an exclusive reaction.  
The density of these final states is then included to calculate the
inclusive differential cross section. The matrix element for the reaction
may be written as

\eq M= \int d\bfr_1 d\bfr_2 d\bfr_3 d\bfr_4  
\psi_{4p}(\bfr_1,\bfr_2,\bfr_3,\bfr_4) <S| F_D
|0> \psi_{He}(\bfr_1,\bfr_2,\bfr_3,\bfr_4)
\qe
where  $\psi_{4p}(\bfr_1,\bfr_2,\bfr_3,\bfr_4) $ is the final state
wave function for 4 protons. We assume that the four protons can be
described by a  product of plane wave states 

\eq 
\psi_{4p}(\bfr_1,\bfr_2,\bfr_3,\bfr_4)=\prod_{i=1}^{4}
e^{i\bfp_i\cdot \bfr_i}
\qe
$\psi_{He}(\bfr)$ is the wave function of the initial state of $^4$He
which is taken here to have a Gaussian form:  
\eq
\psi_{He}(\bfr_1,\bfr_2,\bfr_3,\bfr_4) = N e^{-\alpha
(s_1^2+s_2^2+s_3^2+s_4^2)},
\qe
where $\bfs_i=\bfr_i-\bfR$, $\bfR=(\bfr_1+\bfr_2+\bfr_3+\bfr_4)/4$,
 $N^2=(\frac{\alpha}{\pi})^{\frac{9}{2}}\sqrt{8}$,
$|0>$ is the initial spin state for two neutrons in $^4$He
and $<S|$ is the final spin state of the struck pair. We assume a
box normalization with unit volume.

For the spin independent piece of the pion-nucleon amplitude

\eq 
M_0=
\frac{g(\bfp_3,\bfp_4)}{(2\pi^2)^\frac{3}{2}}
\int d\bfr g(\bfr)e^{i(\frac{\bfk+\bfk'}{2}-\bfp)\cdot\bfr}e^{-\alpha r^2}
\label{eq7}
\qe
where 

\eq
g(\bfp_3,\bfp_4)=8^{\frac{1}{4}} e^{-\frac{(\bfp_3+\bfp_4)^2}{8\alpha}}
e^{-\frac{(\bfp^2_3+\bfp_4^2)}{4\alpha} }
\label{gfac} 
\qe
with $\alpha=\frac{p_f^2}{3}$ and

\eq  
g(\bfr)=\int d\bfq \frac{e^{i\bfq\cdot\bfr}}{k_0^2-q^2}f(\bfq,\bfk') 
f(\bfk,\bfq) 
\qe

The differential cross section may be written as
\eq  
\frac{d\sigma(\bfk,\bfk')}{d\Omega dk'}=                
 \int |M|^2 d\bfp_3 d\bfp_4 d\Omega_p \frac{dp}{dk'} 
\qe 
The integral over the angles of the final relative pp momenta is done by
standard numerical methods while the 6 dimensional integral over $\bfp_3$
and $\bfp_4$ is performed by Monte Carlo by sampling the last factor in 
Eq. \ref{gfac}.

For the  more general case of a fully spin-dependent amplitude, we must
consider amplitudes
\eq
F_{\sigma_{z1},\sigma_{z2}}^{\sigma_{z1}',\sigma_{z2}'}=
<\sigma_{z1}',\sigma_{z2}'|F_D|\sigma_{z1},\sigma_{z2}>
\qe
where $F_D$ is the double scattering operator. The calculation of these
amplitude is given is Appendix A.

\subsection{Final State Interaction}

The FSI is expected to play an important role in the low-energy region of
the final nucleons (when the pion loses a small amount of energy) by
increasing the value of matrix element. In this study, we introduce the
effect of final state interaction among nucleons by taking into account of
the interaction between the two protons in the pair which was produced
from the two charge exchanges on the neutron pair.  We calculate the
correction coming from FSI by replacing the plane wave function for the
active pair (the factor $e^{-i\bfp\cdot\bfr}$ in Eq. \ref{eq7}) 
by $\ph({\bfp})$ which contains an interaction in the s-wave. 
The modified matrix element including the FSI is

\eq 
 M_{FSI}=\frac{g(\bfp_3,\bfp_4)}{(2\pi^2)^\frac{3}{2}}
\int d{\bfr}g(\bfr)
e^{i(\frac{\bfk+\bfk'}{2})\cdot\bfr}\ph(\bfp,\bfr)e^{-\alpha r^2}
\qe 
where

\eq 
\phi(\bfp,\bfr) =4\pi\sum_{\ell,m} i^{\ell}\ph_{\ell}(p,r)Y_{\ell}^{m} 
(\bfr)Y_{\ell}^{m*} (\bfp).
\qe
Since we assume that the final pp interaction is important only in the
relative s-wave 

\eq   
M_{FSI}=\frac{g(\bfp_3,\bfp_4)}{(2\pi^2)^\frac{3}{2}}
\int d{\bfr}g(\bfr)e^{i\frac{(\bfk+\bfk')}{2}}
[\phi_0(p,r)-j_0(p r )]e^{-\alpha r^2} +M_0.
\qe
We approximate $\phi_0(p,r)$ as
\eq
\phi_0(p,r)=\frac{\sin(p r+\delta_0)-\sin(\delta_0) e^{-\alpha_n
r}}{pr},  
\qe 
with $\alpha_n= 2m_{\pi}$. $\delta_0$ is the s-wave phase shift
which is calculated by using the pp version of the (modified) 
Malfliet-Tjon potential\cite{mt,gibbsbook}. 

The strength of the FSI can be expected to have a strong correlation
with the outgoing energy of the pion and produce the greatest effect at
high pion energies.  One can see this qualitatively by considering a DCX
reaction on two isolated neutrons at rest. Since the total momentum of the
recoiling pair of protons is equal to the change in the momentum of the
pion we can write the conservation of energy as

\eq
\omega-\omega'=\sqrt{4(p^2+m^2)+(\bfk-\bfk')^2}-2m. \qe

Thus, for a fixed angle, the momentum, p, is a monotonic function of the
final pion energy and the larger the final pion energy, the smaller is the
relative momentum of the two final nucleons and the stronger the final
state enhancement. For this reason we can expect the inclusion of FSI to
increase the cross section at high outgoing pion energies with little
effect at low pion energies.  Figure \ref{h24.thr} shows the result of the
two-nucleon model with (solid line) and without (dashed line) final state
interactions. It is seen that the strongest effect occurs at high final
pion energy where a significant increase in cross section is observed.

\section{Intranuclear Cascade Code}

To get more a realistic picture of the reaction dynamics, we need to 
include the protons in the initial state in our calculation which we do 
by using an Intranuclear cascade (INC) code in which the pion moves
through the nucleus with probabilities of different kinds of interaction
until it escapes or is absorbed. For an overview see Appendix B. 

Because the nucleons interact with classical potentials in this model,
they can be deeply bound since there is no quantum constraint. To correct
for this effect, these bound nucleons are freed and the binding energy
removed from other particles.  In the present case, bound nucleons were
liberated with their final energy in their center of mass being a fixed
number (20 MeV) equal to about what they would have from their Fermi
motion. This value was varied and little effect was seen. The energy
needed to free the particles was taken from the pion which is probably the
right particle in many, but not all, cases.  This feature is perhaps the
most unrealistic aspect of the current INC code.  This defect could
perhaps be corrected using the method of Wilets et al. \cite{wilets} but
such a correction is beyond the scope of the present work.

\subsection{Final State Interactions}

The INC technique calculates the DCX cross section by following the
classical motion of the pion through the nucleus.  While there are
interactions among all of the nucleons through potentials, as described in
Appendix B, they are not of the quantum mechanical nature that was
included in the two-nucleon model when FSI was treated.  This QM
correction is included in the INC by the use of a weighting factor for
each event.  When a DCX event takes place, all of the coordinates of the
pion and the nucleons are known so that the matrix elements $M_0$ and
$M_{FSI}$ can be evaluated. Since the event includes no QM FSI, one can
correct the probability of its occurrence by including a weight which is
equal to the ratio of the squares of these two matrix elements.

\subsection{Pion Production}

The aim is to model pion production, given a knowledge of pion production
from a nucleon in free space.  The total cross section for pion production
from a free nucleon can be computed from the inelasticity in the phase
shifts\cite{arndt} since, for the moderate energies treated here, the
$(\pi,2\pi )$ reaction is the sole significant contributor to the
inelasticity.

Once it has been decided that a pion production event would take place in
free space (based on the ratio of the pion production cross section to the
total cross section), the distribution of the three-particle final state
(two pions and one nucleon) in energy and angle must be modeled.  To
compute these functions the current relevant data are inadequate, so we
have used two-step models.  One possible model is that in which a $\Delta$
and a pion are produced with the $\Delta$ decaying to produce the second
pion.  This mechanism may indeed play a role.  A second mechanism believed
to be important just above threshold, would produce a ``$\sigma$'' meson
and then have it decay into two pions.  The early data favors this
mechanism, as does the fact that the production of an isospin zero pion
pair is dominant.

Here we use the ``$\sigma$'' mechanism only in the sense that the
pion-nucleon system is converted into a two-pion plus nucleon system
(assumed isotropic in the original $\pi$N center of mass) and then the two
pion subsystem is allowed to decay (again isotropically). To model the
apportionment of energy between the nucleon and the two-pion system we 
assume a distribution of the two-pion invariant mass 
proportional to $Q(Q_{max}-Q)$ where $Q$ is the momentum of one of the
pions in their center of mass. Once these variables have been chosen, the
momentum of the remaining nucleon is known. A Pauli blocking test 
(requiring that the nucleon must not have less energy in the final state
than in the initial state) is applied, resulting in a significant fraction
of the possible production events being disallowed. 

The total pion production cross section calculated at 240 MeV is 116
$\mu$b. Since the production cross section from a single free nucleon is
about 600 $\mu$b, considerably less than the maximum possible cross
section is obtained. This reduction is very likely largely due to the two
protons which ``shield'' the neutrons from which production takes place.

 \subsection{Pauli Blocking }

The effect of Pauli Blocking becomes important at low energy. If there is
little energy loss by the pion the energy of the 4 protons will be small.  
In this case there will be strong cancellation among in the Pauli terms as
was discussed by Ref. \cite{Gibbs1}. Hence, we expect a reduction in the
cross section due the Pauli principle. The effect of Paul blocking is
taken into account in this study by using a weighting factor which
consists of an incoherent summation of the squares of the expressions in
Eqs. \ref{a1} and \ref{a2} (taken from Ref \cite{Gibbs1}).

$$
A_{1}=\frac{1}{\sqrt{24}}[2A(1234)+2A(2134)-
A(1423)-A(1324)-A(24321)-A(2341)
$$

\eq
-A(3214)-A(4213)-A(3124)-A(4123)+2A(3412)+2A(4321)]
\label{a1}\qe

\eq
A_{2}=\frac{1}{\sqrt{8}}[A(1324)-A(1423)+A(2431)-A(2341)-A(3214)+
A(4213)A(3124)-A(4123)]
\label{a2}
\qe
where  

\eq
A(abcd)=
e^{i(\bfp_a\cdot\bfr_1+\bfp_b\cdot\bfr_2+\bfp_c\cdot\bfr_3
+\bfp_d\cdot\bfr_4)}
\qe

The momenta in the first two positions correspond to those
associated with the two nucleons on which the charge exchange occurs.
Tests including this effect showed that it was very small at the energies
considered here and this weight was not normally included in the
calculation.

\section {Discussion}
 
\begin{figure}[p]
\epsfysize=200mm
\epsffile{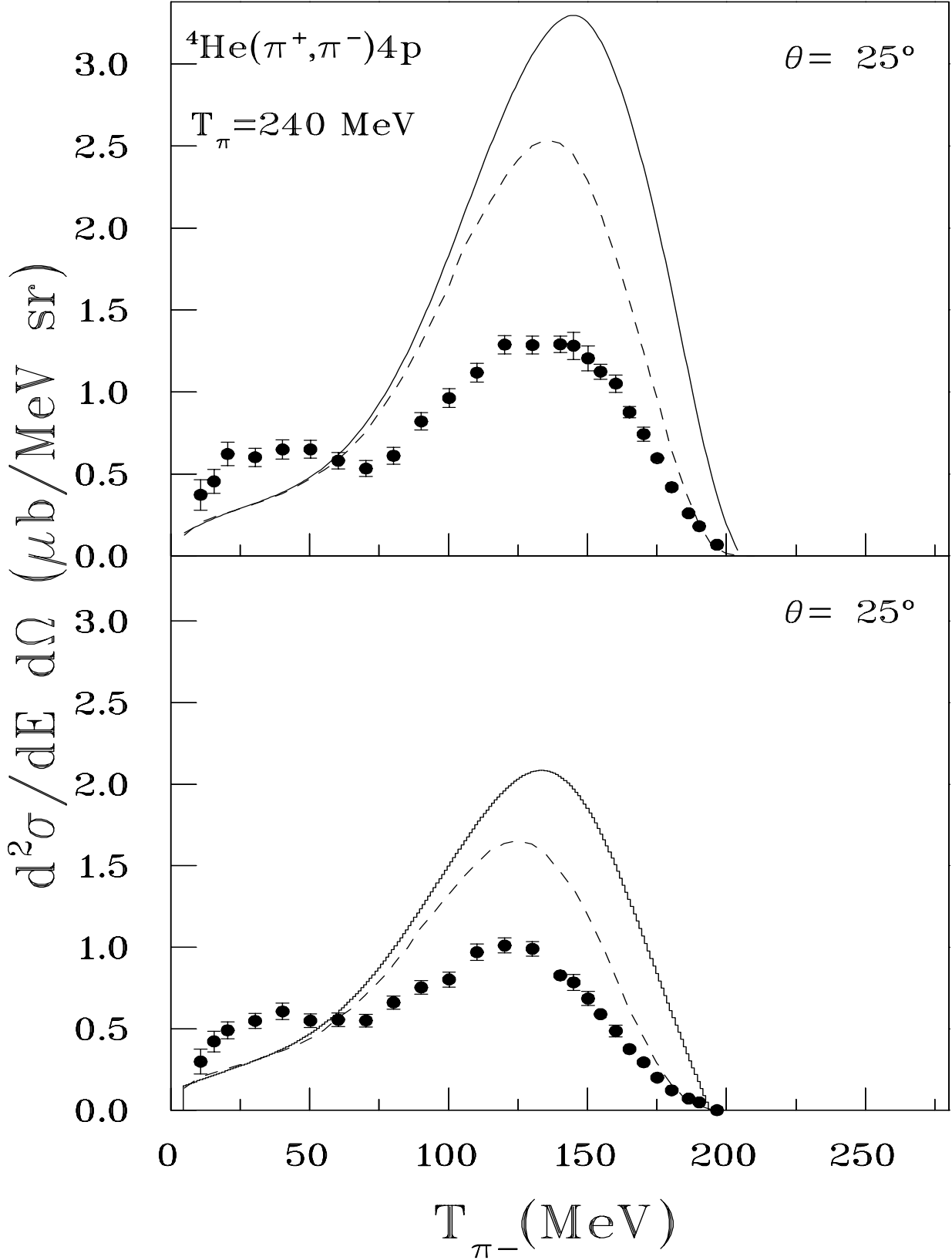}
\caption{Results of the two-nucleon model without (dashed line) and with
(solid line) final state interaction FSI.}
\label{h24.thr}
\end{figure}

We performed calculations for doubly differential cross sections by two
methods, the first being described in Section 2 with the results shown in
Fig. \ref{h24.thr}.  Our results at small angles clearly show the effect
of the FSI which is around 30$\%$ at the peak at 25 degrees.

It is clear from this figure that the two-nucleon model overestimates the
absolute cross section in the region of the high-energy peak.  The
discrepancy could be due to the ignoring of scattering from the two
``spectator'' protons in our calculation. 

\begin{figure}[p]
\epsfysize=200mm
\epsffile{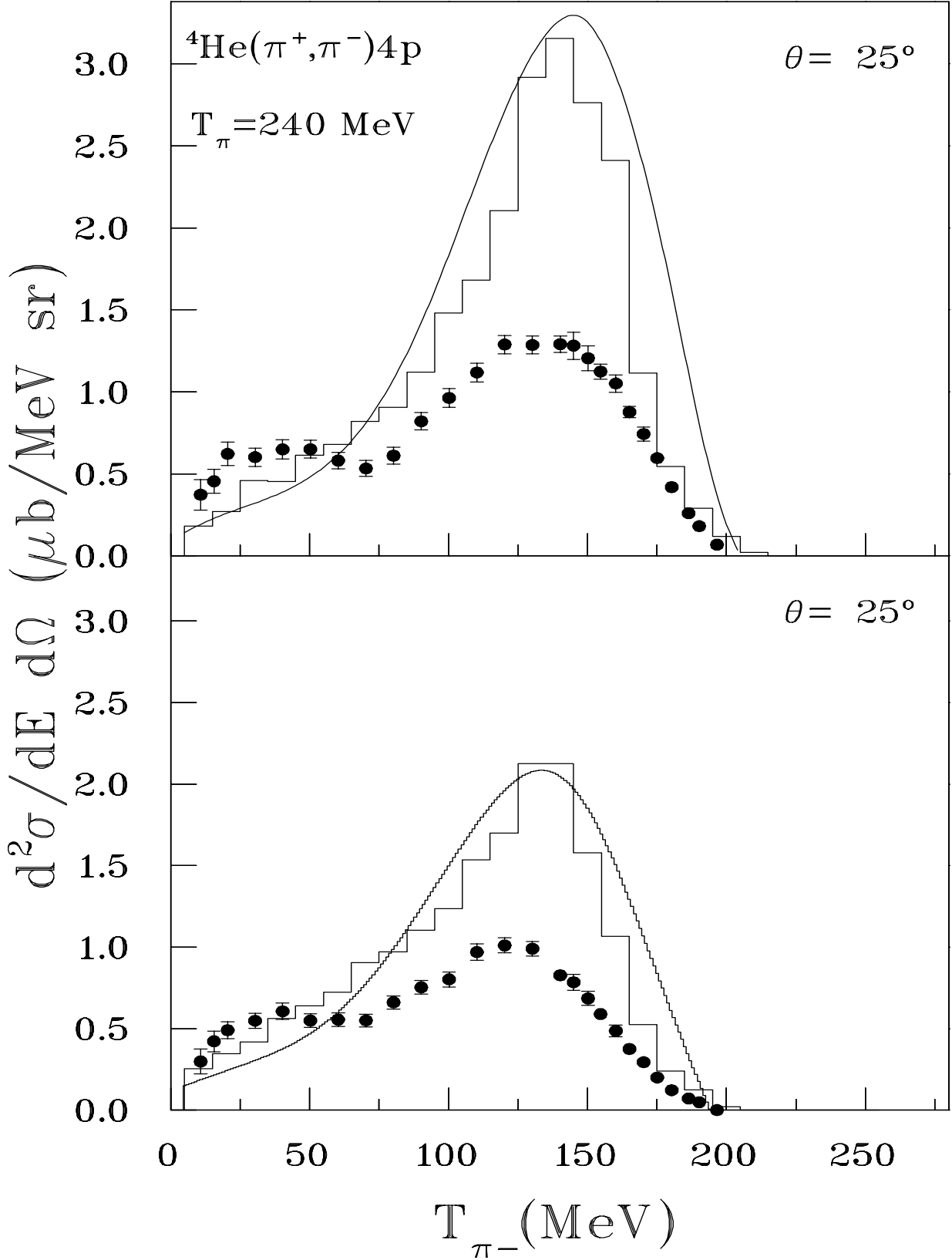}
\caption{Comparison of the INC with only two nucleons (neutrons)
active (histogram) with the two-nucleon model (solid line).}
\label{h24.hthr}
\end{figure}

 Figure \ref{h24.hthr} shows a comparison of the two models under the same
conditions (i.e. only the two neutrons are included in the scattering in
the INC).  The same basic physics is present in the two models (one
classical and one quantum) which are seen to give very similar results at
240 MeV.

\begin{figure}[p]
\epsfysize=200mm
\epsffile{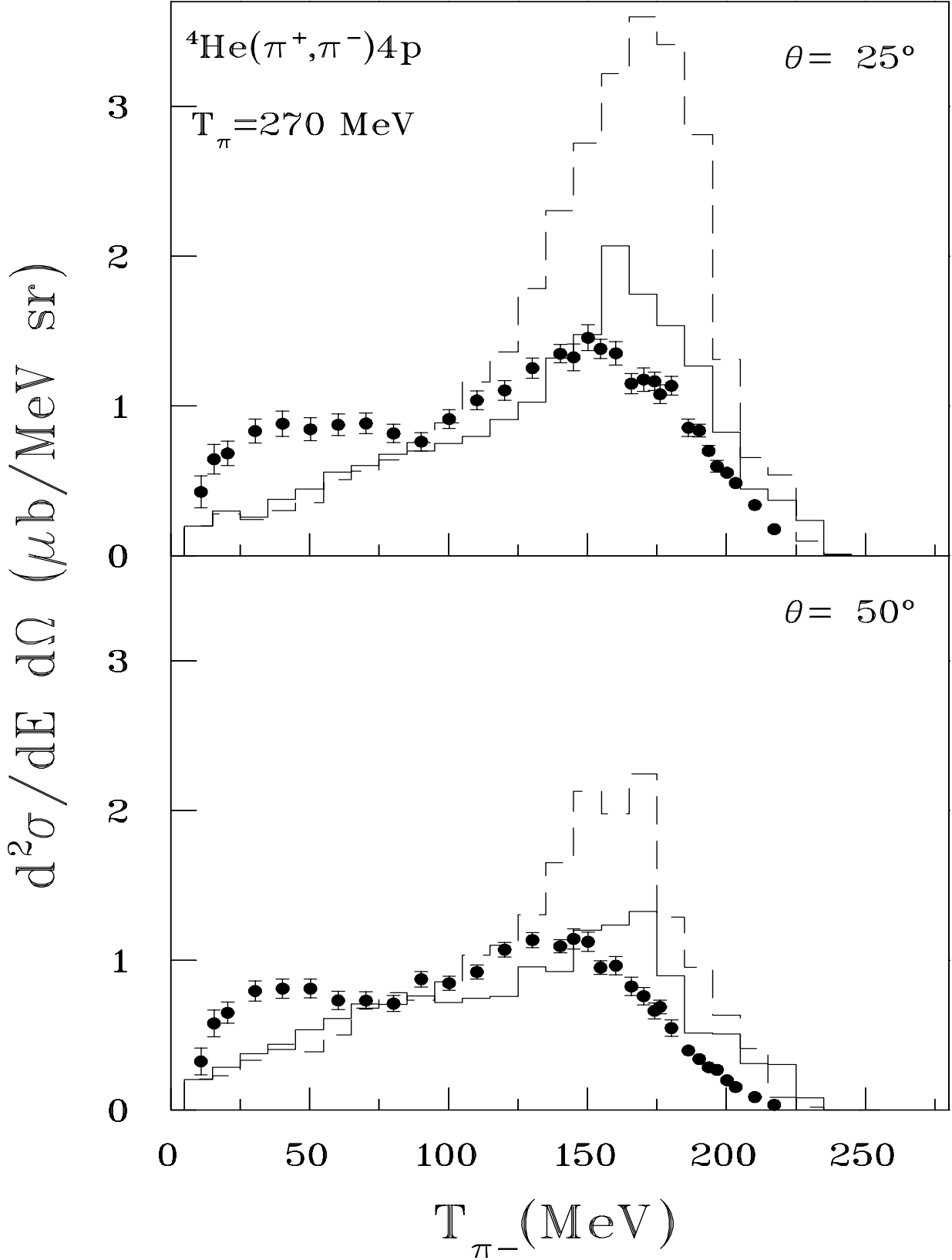}
\caption{The results of INC model at incident energy 270 MeV with
only the  two neutrons (dashed histogram) and four nucleons  (solid
line histogram).}
\label{h27.comp1}
\end{figure}

Figure \ref{h27.comp1} shows the effect of including the two initial
protons at 270 MeV. It is seen that the cross section is considerably
diminished. The two protons are very effective in shielding the neutrons
from the incident pion.

\begin{figure}[p]
\epsfysize=200mm
\epsffile{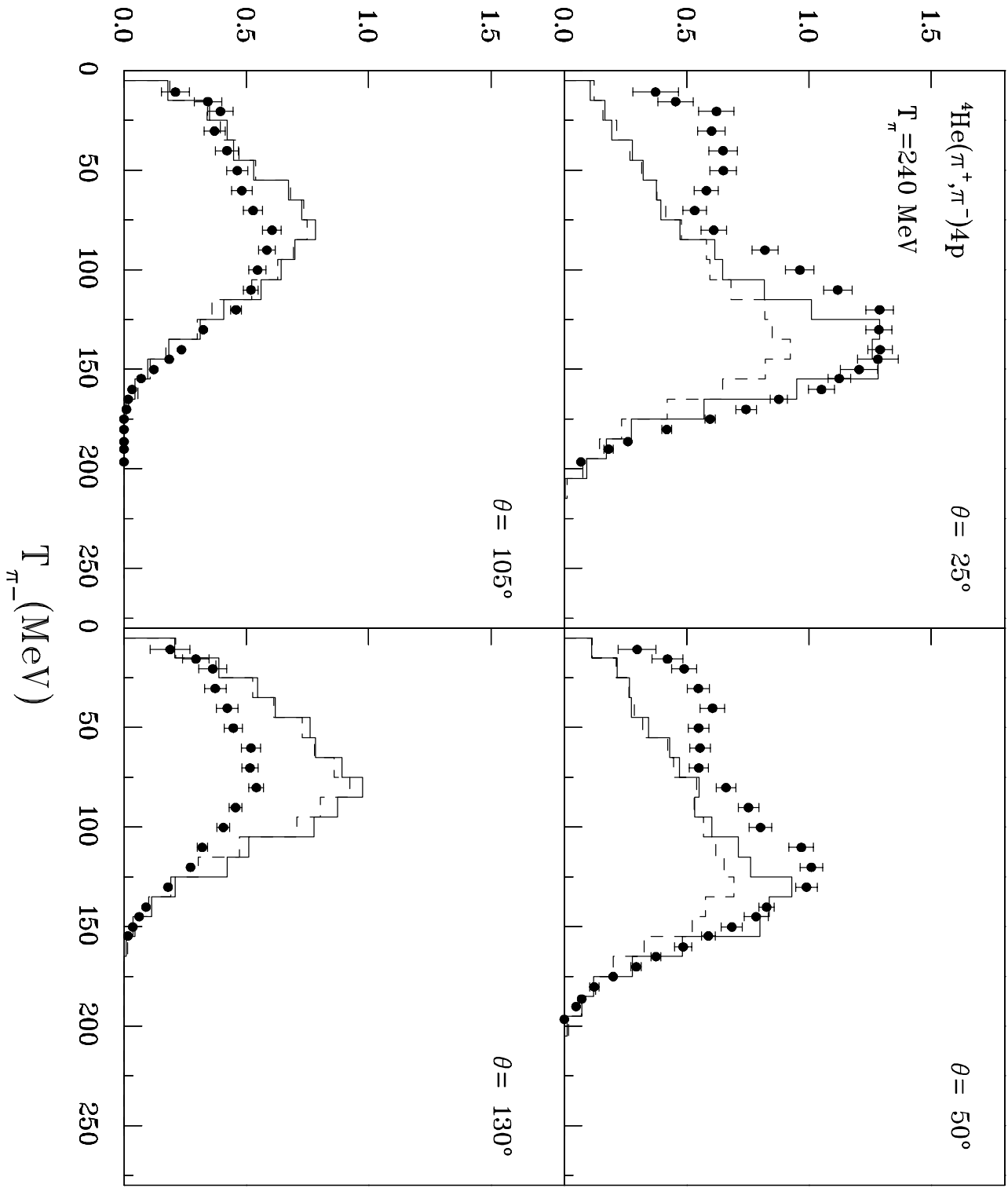}
\caption{The results of doubly differential cross section for the reaction
$^4$He($\pi^+,\pi^-$)4p at incident pion energy of 240 MeV. The
calculation of the solid histogram includes FSI  and the dashed line
histogram does not (absorption is included in both).}
\label{h24ocwabs}
\end{figure}

Figure \ref{h24ocwabs} shows the results with and without the FSI. It is
seen that its inclusion by using a calculated weight in the INC is very
similar to the direct calculation in the two-nucleon model (see Fig.
\ref{h24.thr}), the principal effect being at forward angles and
high-energy outgoing pions.  The shape of the high energy peak is
influenced by the effect of the FSI.

\begin{figure}[p]
\epsfysize=200mm
\epsffile{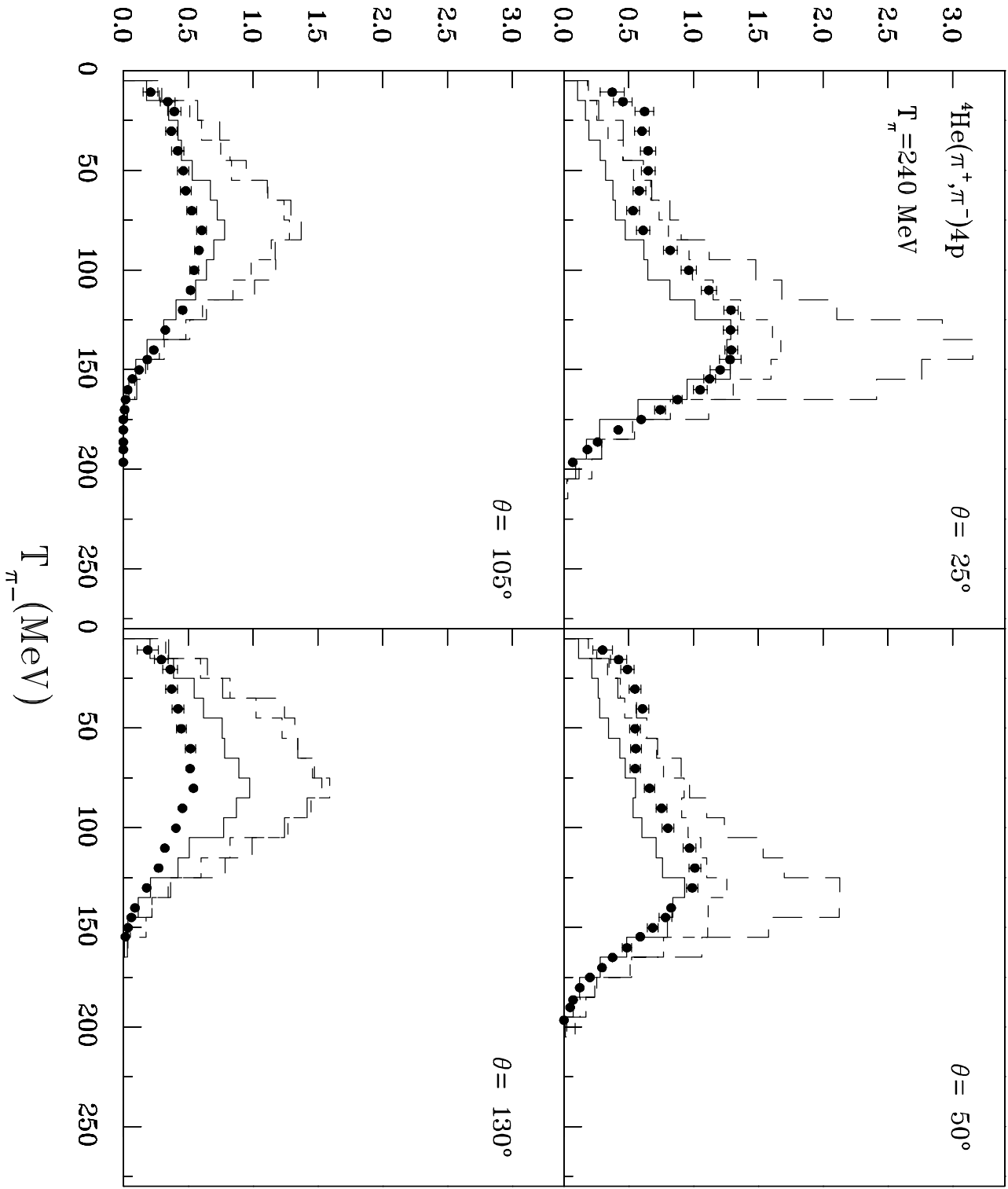}
\caption{The results of the INC at incident pion energy of 240 MeV. The
solid histogram  represents the results including FSI and absorption
factor, the short dashed histogram has no absorption. The long dashed
histogram represents  the results of taking two neutrons only.}
\label{h24.oc.all}
\end{figure}

These effects modify the cross section by up to a factor of 3.  Figure
\ref{h24.oc.all} shows the progression from the case with only 2 nucleons
to the addition of the protons (without true absorption) to the addition
of true absorption.

\begin{figure}[p]
\epsfysize=200mm
\epsffile{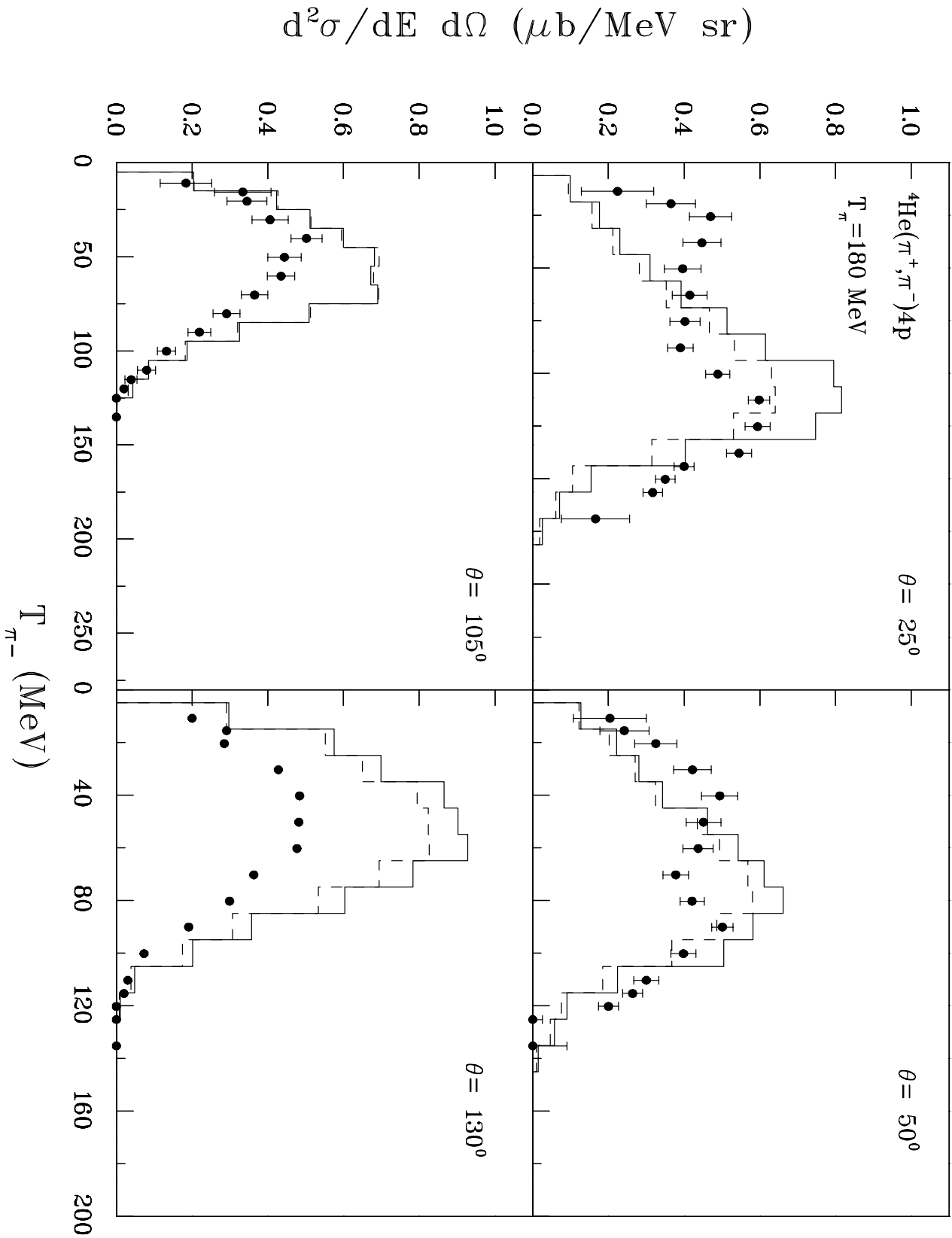}
\caption{The same as Fig. \protect\ref{h24ocwabs}  but at incident  pion
energy 180 MeV.}
\label{h18.wabs1}
\end{figure}

Figure \ref{h18.wabs1} shows the effect of FSI at 180 MeV.  Our
calculation overestimates the cross section but less than the other
calculations \cite {kinney,kul}. That could be due to the fact that 
180 MeV is very close to the $P_{33}$ resonance peak, so that we expect a
great deal of absorption to occur. It is also possible that the INC does
not give good results at the resonance peak because there is a larger
degree of coherence, not taken into account by the classical INC. 

\begin{figure}[p]
\epsfysize=200mm
\epsffile{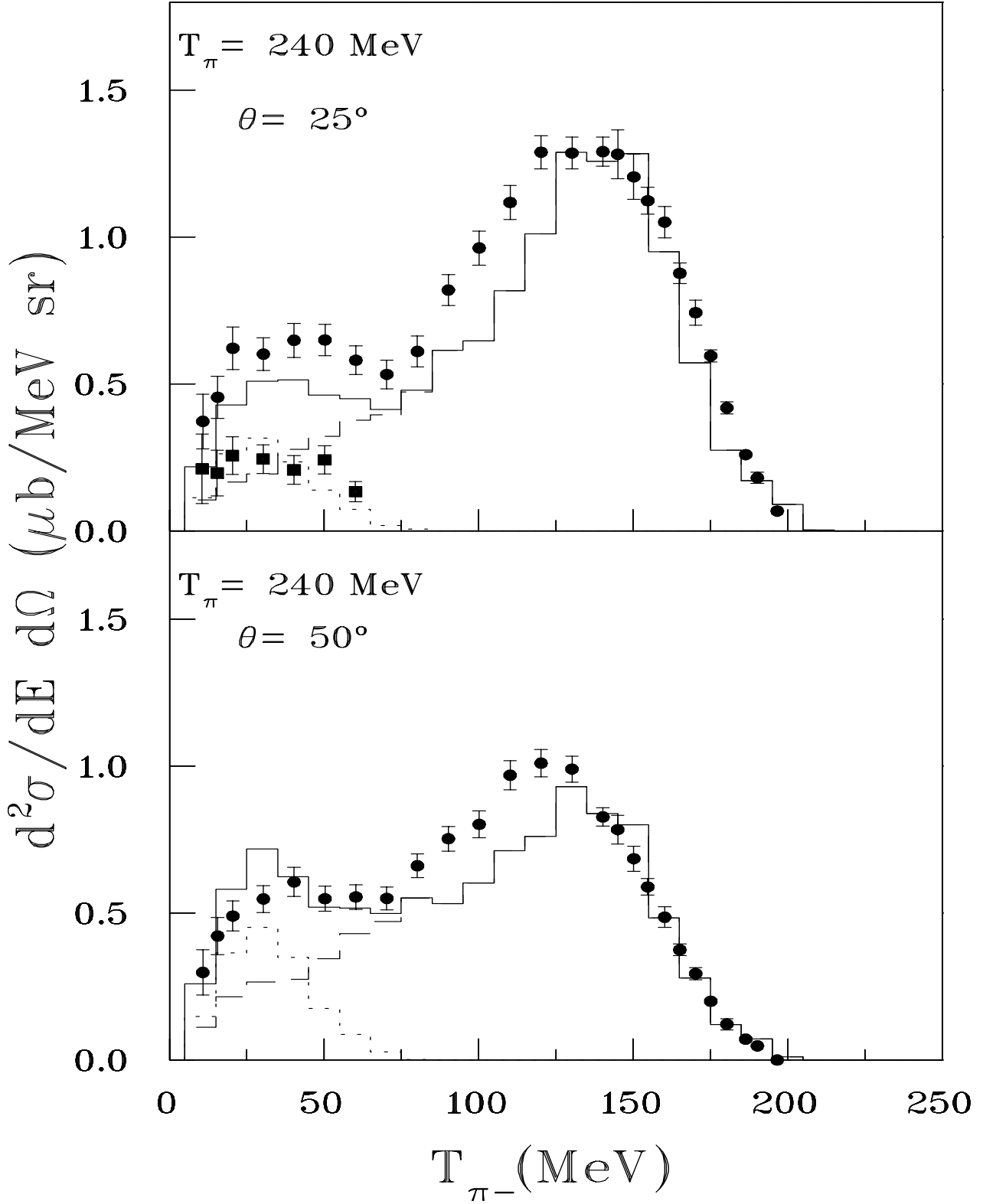}
\caption{The results of doubly differential cross section at 240 MeV. The
solid histogram shows our results with FSI, absorption with pion
production, while the dashed line shows the result  without pion
production. The doted  line shows the pion production cross section
alone. The data given in the solid squares are the production cross 
sections from Ref. \protect\cite{Yuly} multiplied by two.}
\label{h24.oc.pi}
\end{figure}

\begin{figure}[p]
\epsfysize=200mm
\epsffile{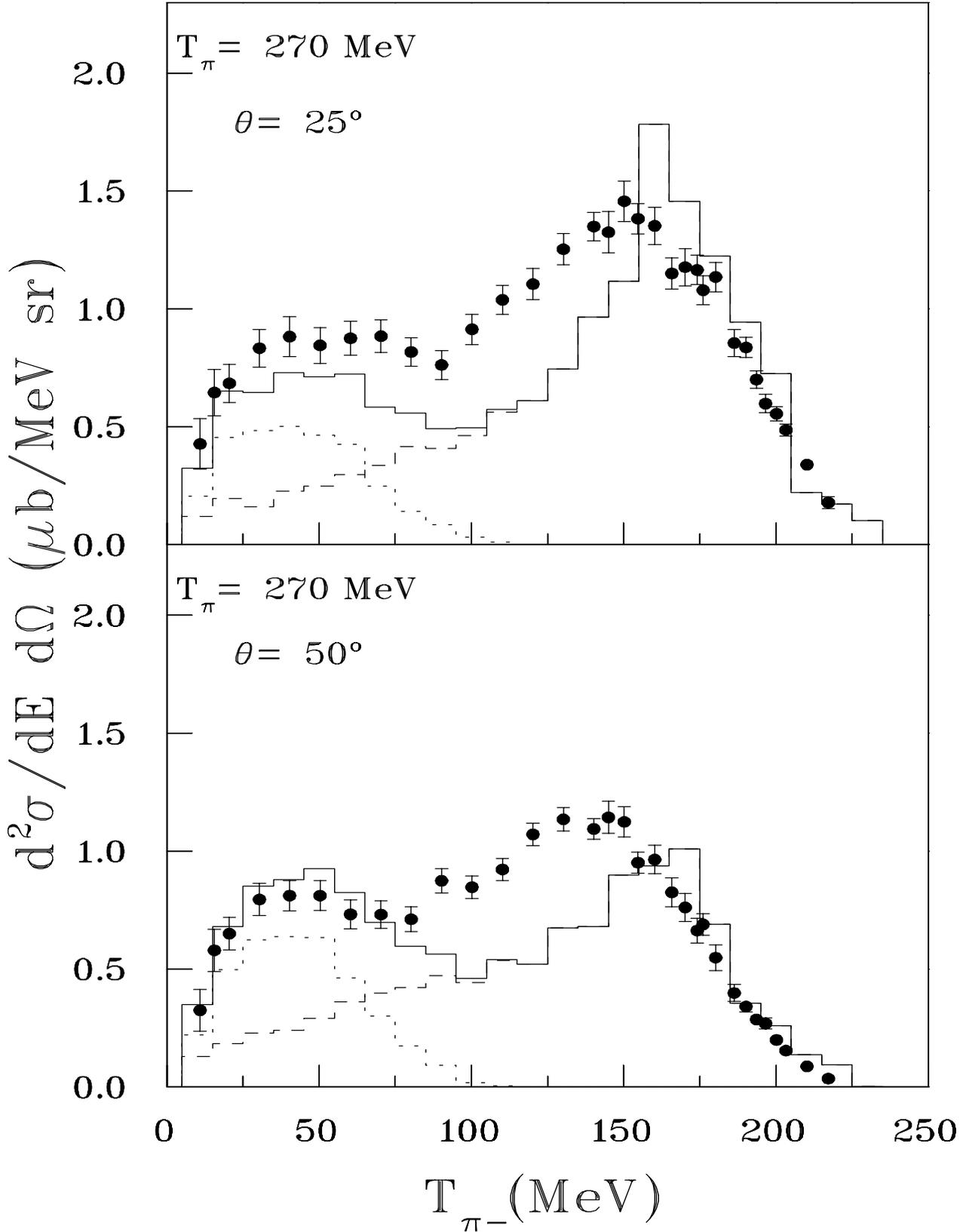}
\caption{The same as Fig. 8 but at an incident energy of 270 MeV.}
\label{h27.pi}
\end{figure}

While the full calculations (solid curves in Figs. \ref{h24ocwabs} and
\ref{h24.oc.all}) give good results for the high-energy peak, the
low-energy peak is underestimated by about a factor of two at forward
angles. Figures \ref{h24.oc.pi} and \ref{h27.pi} show the effect of
inclusion of pion production.  Pion production complements the DCX channel
at pion incident energies of 240 and 270 MeV, and influences the shape and
value of the spectra in the region of the low energy peak.  Vicente \ea
\cite {Vicente} found that the pion production enhances the strength of
the DCX cross section by 15$\%$ at 240 MeV and 30 $\%$ at 270 MeV in
$^{16}$O$(\pi^+,\pi^-)X$. This same mechanism apparently explains the
missing strength at the low energy peak at 270 MeV and 240 MeV in the
present case.

A measurement of pion production from $^3$He was made in Ref. \cite{Yuly}.
In this study negative pions were detected with a incoming \pip\ beam.  
Since there is no double charge exchange possible, the \pim\ mesons must
arise from pion production.  Since there is only a single neutron in
$^3$He, whereas in $^4$He there are two and since the blocking effects of
the two protons should be about the same in the two cases, we expect the
production in $^4$He to be about twice that in $^3$He.  In Fig.
\ref{h24.oc.pi} we show the data from Ref. \cite{Yuly} multiplied by a
factor of two.  The agreement with this data under this assumption is
satisfactory.

In regard to the possible forward-backward explanation of the two peaks,
we note first that the effect is being observed in the laboratory. Even if
the single charge exchange angular distribution remained forward-backward
symmetric in the center of mass, it would become more forward peaked in
the lab as the energy increases which would lead to the high energy peak
becoming increasingly dominant. In fact the single charge exchange angular
distribution becomes more forward peaked in the center of mass as well.
These two effects lead one to expect a decreasing low-energy peak with
rising beam energy since it is supposed to arise from the back scattering
part of the angular distribution.  The data show the opposite effect with
the low-energy peak growing with respect to the high-energy one with
increasing energy.  This observation is naturally explained if pion
production is important since its contribution rises with energy.

\begin{figure}[p]
\epsfysize=200mm
\epsffile{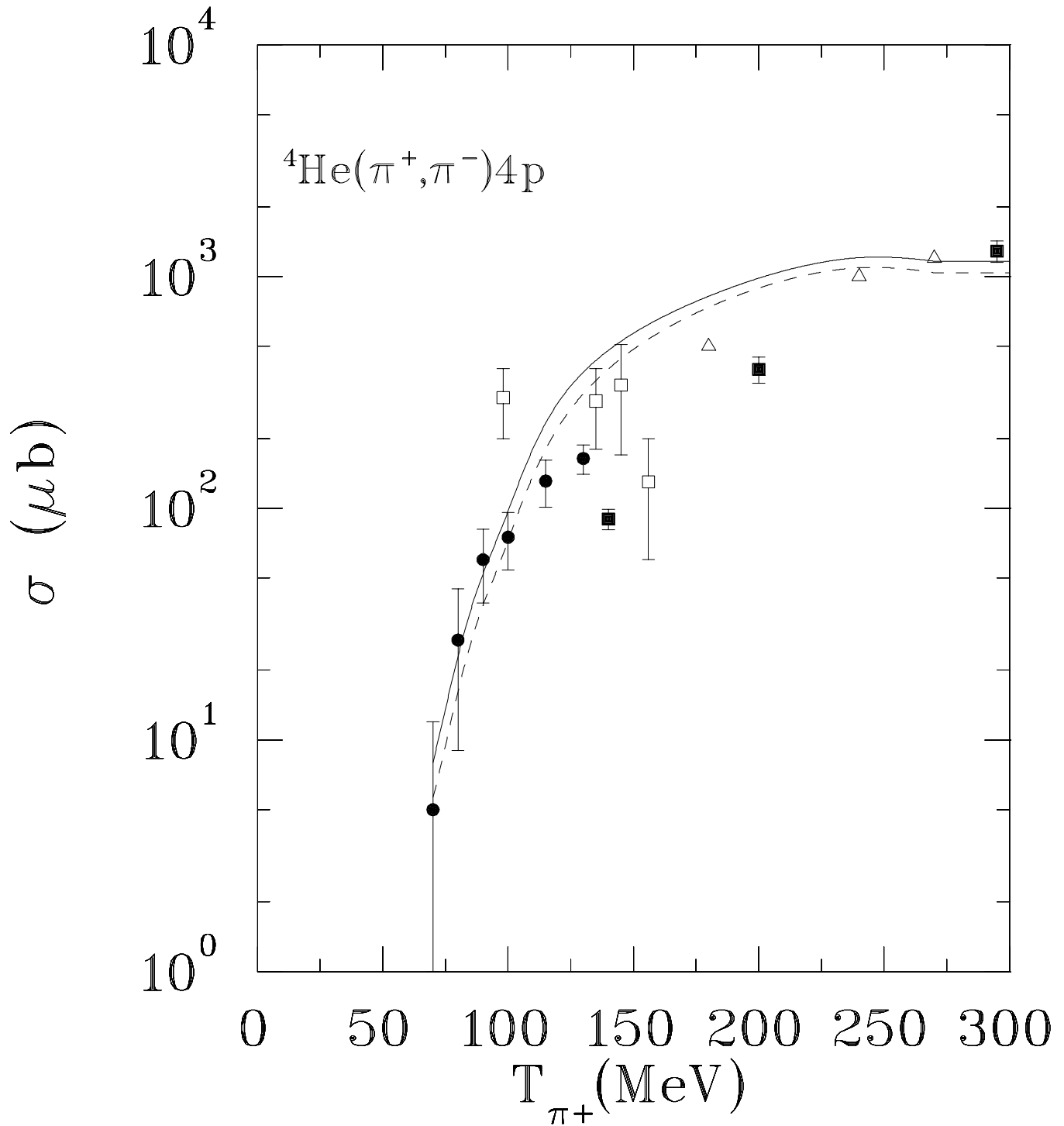}
\caption{Total DCX cross section for $^4$He($\pi^+$, $\pi^-$)4p. The
solid line shows the results including FSI and absorption factors,
while the dashed line is without FSI.  The experimental data were
taken from \protect{\cite{Bilger}} (solid circles), 
\protect{\cite{kinney}} (open triangles), \protect{\cite{falomkin}} (open
squares) and \protect{\cite{stetz}} (solid squares). 
}
\label{dcx.tabs}
\end{figure}

Figure \ref{dcx.tabs} shows the integrated total DCX cross section
with the INC compared with data.

\section {Conclusion}

In this paper, we have presented calculations of the double charge
exchange cross section for reaction $^4$He $(\pi^+, \pi^-) 2$p by using a
sequential mechanism and an INC model. The consideration of FSI in our
calculation enhances the high energy peak at small angles by 30 $\%$
leading to good agreement with experimental data. The missing cross
section strength in the low energy peak in our calculation at energies 240
and 270 MeV is naturally explained by pion production.

We note that there is a discrepancy between theory and experiment in
regard to the shape of the 180 MeV data.  While inclusive pion production
gives only a very small contribution (one is only about 5 MeV above
threshold) a significant contribution can come from the production leading
to $^3$He plus a proton. The threshold for this reaction is about 167 MeV. 
The calculation of this cross section is beyond the scope of the
techniques used in the present paper. 

We wish to thank Mark Yuly for supplying the numerical values for the
production cross section on $^3$He from Ref. \cite{Yuly}. This work
was supported by the National Science Foundation and the Jordan 
University of Science and Technology.

\begin{appendix}

\section{Spin Dependence}

To include spin in the calculation, it is preferable to rewrite the 
pion-nucleon scattering amplitude in a slightly different form. 

By using the relations

\eq 
{\bf {\hat q}}\cdot{\bf {\hat q'}}=-\frac{4\pi}{\sqrt{3}}\sum_{m,m'} 
C_{1,1,0}^{m,m',0}Y_1^{m}({\bf {\hat q}})Y_1^{m'}({\bf {\hat q'}}) 
\qe
and

\eq{\bf {\sigma}}\cdot{\bf {\hat q}}\times{\bf {\hat q'}}
=\frac{8\pi i}{3\sqrt{2}} \sum_{m,m',M} C_{1,1,1}^{m,m',M}\sigma^{-M}
Y_1^{m}({\bf {\hat q}})Y_1^{m'}({\bf {\hat q'}}) \qe

we can then express the $\pi N $ amplitude as  

\eq     
\bf{f(q,q')} = \sum_{\beta=0,1,2}(\frac{1}{\sqrt{3}})^{\gamma}
D_{\beta}(i)q^{\alpha}q'^{\alpha}v_{\alpha}(q)v_{\alpha}(q')
\sum_{m,m',M}C^{m,m',M}_{\alpha,\alpha,\gamma}t^{-M}_{\gamma}(i)
Y^m_{\alpha}({\bf{\hat q}})
Y^{m'}_{\alpha}({\bf{\hat q'}}) 
\qe
where $\alpha$,\ $\beta$ and $\gamma$ are related by the following table.

\begin{center}
\begin{tabular}{|c|c|c|} \hline
$\beta$ & $\alpha$ & $\gamma$ \\ \hline
0 & 0 & 0 \\ \hline
1 & 1 & 0 \\ \hline
2 & 1 & 1 \\ \hline
\end{tabular}
\end{center}

The matrices $t_{0}$ and $t_{1}^{\mu}$ are defined by

$$ 
t^0_0 = \left( \begin{array}{cc}
1 & 0 \\
0 & 1
\end{array}
\right) 
$$

$$ 
t^0_1=\sigma^0=\sigma_z;\ t^1_1=\sigma^1=-\frac{1}{\sqrt{2}}
(\sigma_x+i\sigma_y);\ t^{-1}_1=\sigma^{-1}=\frac{1}{\sqrt{2}}
(\sigma_x-i\sigma_y) 
$$
and

$$ 
D_{0}(i)=4\pi\lambda_{0}(i);\ D_{1}(i)=\frac{4\pi}{\sqrt{3}}
\lambda_{1}(i); D_{2}(i)=\frac{8\pi i}{\sqrt{6}}\lambda_{2}(i) 
$$

The operator in spin space for the double scattering will then be given by

\eq
{\bf F}_{D}({\bf k},{\bf k'})=\sum_{\beta_{1},\beta_{2}} D_{\beta_1}(1)
D_{\beta_2}(2) k^{\alpha_1} k'^{\alpha_2} Y^{m_1}_{\alpha_1}({\bf{\hat k}}_1)
Y^{m'_2}_{\alpha_2}({\bf{\hat k}}_2)\times  
\qe

$$ 
( \frac{1}{\sqrt{3}} ) ^{\gamma_1+\gamma_2}
C^{m_1,m'_1,M_1}_{\alpha_1,\alpha_1,\gamma_1}
C^{m_2,m'_2,M_2}_{\alpha_2,\alpha_2,\gamma_2} 
t^{-M_1}_{\gamma_1}(1) t^{-M_2}_{\gamma_2}(2) 
$$

$$ 
\times A^{m'_1,m_2}_{\alpha_1,\alpha_2}({\bf k}_{0},{\bf r})
$$
where

$$
 A^{m'_{1}m_{2}}_{\alpha_{1}\alpha_{2}}(k_{0},r)=\int
q^{\alpha_1+\alpha_2}
Y_{\alpha_1}^{m_1}({\bf {\hat q}})Y_{\alpha_2}^{m_{2}^{*}}({\bf {\hat
q}})
{\left( \frac{\alpha^2 + k_{0}^2}{\alpha^2 + q^2}\right)}^2
\frac{e^{i{\bf q}\cdot{\bf r}}}{(k^2_{0}-q^2)}
d{\bf q} $$                                    

We are now ready to calculate the matrix elements of this operator in spin
space. Using  

 \eq <\sigma'|t^{-M}_{1}|\sigma>=-\sqrt{3}C^{\sigma,-M,\sigma'}_
{\frac{1}{2},1,\frac{1}{2}} \qe
we can write  

$$ <\sigma'_1,\sigma'_2|{\bf F_{D}}({\bf k},{\bf k'})|
\sigma_1,\sigma_2>\equiv{\bf F_{D}}^{\sigma'_1,\sigma'_2}
_{\sigma_1,\sigma_2}
({\bf k},{\bf k'}) $$

$$
 = \sum_{\beta_1,\beta_2}(-1)^{M_1+M_2}
D_{\beta_1}(1)D_{\beta_2}(2){\bf k}^{\alpha_1}{\bf k'}^{\alpha_2}
Y^{m_1}_{\alpha_1}({\bf{\hat k}})
Y^{m'_2}_{\alpha_2}({\bf{\hat k'}}) \times 
$$
\eq
C^{m_1,m'_1,M_1}_{\alpha_1,\alpha_1,\gamma_1}
C^{m_2,m'_2,M_2}_{\alpha_2,\alpha_2,\gamma_2} 
C^{\sigma_1,-M_1,\sigma'_1}_{\frac{1}{2},\gamma_1,\frac{1}{2}}
C^{\sigma_2,-M_2,\sigma'_2}_{\frac{1}{2},\gamma_2,\frac{1}{2}}  
A^{m'_1,m_2}_{\alpha_1,\alpha_2}({\bf k}_{0},{\bf r})
\qe

\section{The Intranuclear Cascade}

The present code was originally developed to treat moderate-energy
antiproton annihilation in nuclei and has been applied to that end several
times \cite{rice,wrg,gkruk,fermilab,gs,anti1}. The annihilation of an
antiproton leads to pions (or at least it is so treated by the code) and
so the history of pions in the energy range below and of the order of 1
GeV are essential to the calculation of energy deposition.  It has had
considerable success in predicting the rapidity distributions of strange
particles produced in antiproton reactions\cite{gkruk}.  The question of
pion absorption and comparison with data has been addressed \cite{lanl}
and the code has been used for the comparison with inclusive data in 
work by Zumbro \ea\cite{zumbro}.

The initial bombarding projectile is started with appropriate initial
momentum toward a circle which is large enough to contain the
projected density of the target nucleus.  A fraction of the beam particles 
(usually about one half) pass without interacting and cross sections
are computed as the fraction of the reactions of interest which
occur multiplied by the area of this disk. 

\subsection{Treatment of Nucleons}

The heart of the code is the construction of a model nuclear system 
with as many characteristics of a ``real'' nucleus as possible.  The 
code exists in several versions according to the reaction being
investigated and the detail required.

\subsubsection{Version I: The Woods-Saxon Potential}
In this original model the nucleon-nucleon interaction is thought 
of as being made up of two parts: 1) a long range interaction (one 
pion exchange) and 2) a short-range part (heavy meson exchange, quark 
interaction, etc.).

The average of the long-range part is represented by a potential with 
a Woods-Saxon shape.  The parameters of the well are chosen so that the 
particle distribution that results from motion in this well represents 
the measured proton densities in nuclei.

Each nucleon has a designated binding energy specified in the input data. 
For a given binding energy there is a point at the edge of the well where
the potential is equal to the binding energy and hence the kinetic energy
is zero.  For each particle a limiting radius, $R_x$, is chosen as that
value of the radius and a set of position coordinates are chosen uniformly
distributed within a sphere of radius $R_x$.  Once the position for a
given nucleon is established, its potential energy can be calculated and,
using the specified binding energy, the kinetic energy and hence the
magnitude of the momentum, can be obtained.  The directions of the momenta
are chosen randomly in an isotropic manner.  Since each particle is moving
in a conservative potential well it will maintain the same total (binding)
energy throughout its motion unless disturbed by an outside agent.

The short-range part of the nucleon-nucleon interaction is represented by
a scattering cross section.  

\subsubsection{Version II: Inter-nucleon Potentials with Fixed Well}

In this version the potential well is generated by the superposition of
semi-realistic nucleon-nucleon potentials due to the nucleons themselves,
those by Malfliet and Tjon \cite{mt,gibbsbook}. 

If $V_i$ is the potential seen by the $i^{th}$ nucleon at position
$\bfr_i$ then

\eq V_i =\h\sum_{i\ne j}V_{ij}(|\bfr_i-\bfr_j|) \qe
where the subscript $ij$ on $V$ is to label whether the interaction is np
or pp (nn). For the technique of creating the initial nucleus see section
\ref{b3}.

If we wish to guarantee that the nucleus remains in the ground state 
(i.e. with each nucleon keeping its assigned binding energy) 
we can fix the wells to those generated by the originally chosen 
nucleons with each nucleon  moving in its own fixed well, even though 
the other nucleons which created this potential have moved to other 
positions as the time evolution progresses.  Since the movement
of non-struck nucleons is small, the actual wells should not 
be very different from these fixed wells. The reason for wanting to 
keep the fixed wells is to maintain the quantum condition that the 
nucleus remain intact with the same ``quantized'' states unless acted 
upon by an external force. As before, the nucleons can collide with a 
probability computed from a cross section entered as data.  Note that 
this is necessary since the nucleons cannot exchange energy except by 
this mechanism.  A nucleon may scatter from another nucleon's initial 
position but it simply bounces off of the potential and the
nucleon ``struck'' knows nothing of the interaction.

One inconvenient point which is especially bothersome for light nuclei is
that a single nucleon can be left bound in the phantom potential well. For
example for $^4$He if, in $\pi^+$ absorption, three protons are removed,
the remaining neutron is still bound in the well (if it is not
specifically struck by the incident pion or one of the leaving nucleons) 
and is left with a negative energy when it should have a positive energy
distribution (presumably related to its Fermi momentum). 

It is this version which was used for the inclusive measurements by Zumbro
\ea\ \cite{zumbro}. 

\subsubsection{Version III: Inter-nucleon Potentials with a Changing Well}

In this case the calculation of the potential wells follow the nucleons as
they move.  In some sense this is the most realistic of all the models. It
has problems, however.  As in any classical model the particles can
evaporate from the system.  If the nucleus falls apart rapidly the density
at which the reaction takes place may not be correct, even though the
initial nucleus is originally thrown with an acceptable density.  To
minimize this problem the nucleons do not begin to move until there is a
first interaction of the projectile with one of them.  Note that a typical
 nuclear reaction takes place on the order of a few fm/c.  Thus, while the
nucleus is decreasing in density as a function of time, the problem is not
very serious.

\subsection{Interaction of the Projectile with the Nucleons}

The next step is to allow the projectile to propagate through the nucleus
and interact with the nucleons.  Only pions will be treated in the present
discussion.  Kaons, $\eta$'s and other mesons are also propagated by
similar techniques. Immediately after moving the nucleons one time step
the meson multiple scattering subroutine is entered.  This subroutine
moves the pion (or pions) one time step, taking into account the
possibility that it can elastically scatter, charge exchange, be absorbed
on a nucleon pair or produce one or more additional pions. 

For each pion the following steps are carried out.

The distance from the pion to each nucleon is computed and compared with
the value of R derived from the equation

\eq \sigma_{max}=\pi R^2 \qe
where $\sigma_{max}$ is the maximum pion-nucleon total cross section (that
at the 33 resonance).  If the distance is greater than R the index of the
loop passes on to the next nucleon.  The total cross sections used for
comparison are $\pi^0$ cross sections, or in other words, the average of
$\pi^+$ and $\pi^-$ cross sections.  A weight is carried for the event
which is the ratio of the true total cross section to the $\pi^0$ total. 

If the distance is less than R then the possibility that an interaction
might take place is pursued.  From the momenta of the pion and the nucleon
the effective laboratory momentum for the pion is computed, i.e. the value
the pion would have (for the same center of mass energy) if the nucleon
were at rest.  This is done because c.m. cross sections are tabulated as a
function of $P_{lab}$.  The distance from the pion to the nucleon is then
compared with a length, R, obtained from $\sigma_T(P_{lab})=\pi R^2$ where
$\sigma_T$ is tabulated in a vector indexed by $P_{lab}$.  If the distance
is greater than this number then there is no interaction with that
nucleon. 

When a case is found that the distance to the nucleon is less than
$R$ then a weight for the event is computed from the ratio of cross
sections and one of a series of branchings is selected according to
conservation laws and the appropriate probability.

First, the possibility of pion absorption is explored.  In order for a
pion to be absorbed it must have interacted previously with at least one
nucleon and absorption must be possible (charge conservation). If those
two conditions are met, then a random number is compared with an input
parameter, $P_{abs}$, to determine if an absorption will take place. 

If the pion is absorbed then the total energy of the pion is converted
into momentum added to the current and previous nucleon in such a way that
the sum of the additional momenta are zero (they are equal and opposite)
and along the direction of the trajectory of the pion traveling between
the two nucleons, i.e. parallel to the current pion momentum.  If $\ppi$
is the current value of the pion momentum then the current nucleon will
receive

\eq \bfP +\frac{\ppi}{2}\equiv \alpha\ppi \qe
and the previous nucleon
\eq -\bfP +\frac{\ppi}{2} \equiv \beta\ppi. \qe
To conserve energy we have (non-relativistically)
\eq \frac{(\bfP_1+\alpha\ppi )^2}{2m} +\frac{(\bfP_2+\beta\ppi )^2}{2m}
=\frac{P_1^2}{2m}+\frac{P_2^2}{2m}+\omega \qe

Where 1 (say) is associated with the current nucleon and 2 with the
previous one.  Solving for $\alpha$ and $\beta$ the momentum $\bfP$ can be
found.  The direction of $\ppi$ relative to the incident beam direction
(in the case of only two nucleons) approximately follows a p-wave
distribution from the previous elastic scattering. The direction of this
large momentum dominates the angular distribution and tends to give a
reasonably good representation of $\pi d$ absorption. 

If pion absorption does not take place then the possibilities of pion
production (if energy conservation permits), charge exchange (if charge
conservation permits) or elastic scattering are chosen according to the
ratio of the relevant cross sections to the total.

If pion production does not take place then charge exchange or elastic
scattering occur.  In this case a table of Legendre polynomial
coefficients are used (from a data table) to select an angular
distribution for the final two particles.  If the scattering or charge
exchange leads to a nucleon whose energy is not greater than its initial
value by the amount of the Pauli blocking energy the scattering is not
allowed. 

\subsection{Technique for creating a Nucleus with correlations
\label{b3}}

The procedure starts from the point of view of the shell model. The
 full Hamiltonian can be written as

\eq \sum_{i}T_i+\frac{1}{2}\sum_{j\ne i}V_{ij}(\bfr_i-\bfr_j)=E \qe

This equation can be regarded as a classical or operator relationship.

In the spirit of the shell model it is natural to write the total
energy as the sum of single-particle energies $ E=\sum_{i}E_i $
and make the association

\eq T_i+\frac{1}{2}\sum_jV_{ij}(\bfr_i-\bfr_j)=E_i \qe for each particle. 
Certainly this is not necessary but it is sufficient to satisfy the
original equation and it allows us to associate an energy with each
particle.  These equations may be regarded as operator equations for the
Schr\"{o}dinger equation or simply classical energy conservation
equations.  A significant difference is that in the quantum case the
kinetic energy symbols mean derivative operators and in the classical case
they are non-negative quantities.  In the quantum case they lead to
negative multipliers of the wave function only in the case of tunneling. 
Since tunneling is a small probability we make the approximation that the
kinetic energies should always be positive. In any event, we are forced to
require that the kinetic energies be non-negative for the construction of
the nucleus since it is to be used in a classical simulation. 

The resulting algorithm can be described as follows:

1) All of the particles are assigned a value of binding energy in 
the shell model sense.  As an example, for $^{12}$C, there might be 
4 particles with a binding energy suitable for the s-shell and 8 
particles with a binding energy in the p-shell. The sum of these 
binding energies should be equal to the total binding energy of the 
nucleus.

2) Each shell is assigned a sphere with some defined radius.  This 
radius parameter allows a fine tuning of the radius for each shell 
to be obtained.  If this radius is taken too small the nucleus will 
be compressed.  If it is taken too large time will be wasted in 
attempting to create impossible configurations.  In general it 
should be taken of the order of the effective square well radius of 
a given shell.

3) A nucleon-nucleon potential is  assigned to each particle pair. In 
general this is defined between particle types (nn and pp or np). 

4) The actual creation of the nucleus is begun with the throwing of 
a random uniform distribution of particles within the spheres defined 
above.

5) The calculation of each kinetic energy is made using the positions 
of the nucleons to calculate the potential energy and subtracting it
from the individual binding energies.  If any kinetic energy is 
negative then the complete re-throw of the nucleus is started again with 
step 4.  If all kinetic energies are non-negative then the proposed
nucleus 
is possibly viable.  Note that the particles cannot be very close or 
the repulsive potential will make some kinetic energy negative nor can 
they be too far apart for the same reason.

6) Now the magnitudes of the momenta are calculated from the kinetic
energies and the directions are thrown in a random uniform manner.  
This process will lead to a nucleus with a non-zero total momentum.

7) The total momentum of the nucleus is calculated by vectorially 
adding the individual momenta.  If the magnitude exceeds a set 
limit, the angles of the momentum vectors are thrown again in an
attempt to make the sum less than the limit returning to step
6.  Since it is possible that no combination of angles will lead 
to a total momentum less than the prescribed limit, after some
fixed number of  tries (typically 300) the entire nucleus
is re-thrown by returning to step 4.

8) 1/A of the total momentum is subtracted from each nucleon thus putting 
the nucleus at rest. This procedure does not leave the 
binding energy unchanged but if the limit on the total momentum is 
small the error on the binding energy is also small.  For example,
it was found that if the total momentum is restricted in step 7 to be 
less than 40 MeV/c then the maximum error in the binding energy was 
observed to be less than 0.2 MeV (out of 28 MeV) for $^4$He.

\end{appendix}

\newpage

\end{document}